\documentclass[12pt]{article}
\usepackage{amsmath,amsfonts}

\makeatletter\@addtoreset{equation}{section}\makeatother

\def\bZ {\mathbb{Z}}
\def\bC {\mathbb{C}}

\input epsf

\newcommand{\preprint}[1]{\begin{table}[t]  
             \begin{flushright}               
             {#1}                             
             \end{flushright}                 
             \end{table}}                     
\renewcommand{\title}[1]{\vbox{\center\LARGE{#1}}\vspace{5mm}}
\renewcommand{\author}[1]{\vbox{\center#1}\vspace{5mm}}
\newcommand{\address}[1]{\vbox{\center\em#1}}
\newcommand{\email}[1]{\vbox{\center\tt#1}\vspace{5mm}}

\begin{document}

\begin{titlepage}
\preprint{hep-th/0602103 \\ 
ITFA-2006-06}

\title{On the critical points of the entropic principle}

\author{Bartomeu Fiol}

\address{Institute for Theoretical Physics, University of Amsterdam\\
1018 XE Amsterdam, The Netherlands}

\email{bfiol@science.uva.nl}

\abstract{
In a recent paper, hep-th/0509109, Gukov et al. introduced an entropy 
functional on the moduli space of Calabi-Yau compactifications. The maxima
of this functional are then interpreted as 'preferred' Calabi-Yau 
compactifications. In this note we show that for compact Calabi-Yaus, all 
regular critical points of this entropic principle are maxima.}

\end{titlepage}

\section{Are some vacua more equal than others?}
A feature of M/string theory that is receiving increasing attention is the 
vast number of vacua, even if we decide to restrict ourselves to static 
solutions with 3+1 macroscopic dimensions and ${\cal N}=0,1$ (the papers 
devoted to the subject constitute a landscape on their own; see {\it e.g.} 
\cite{Douglas:2004zg} for reviews). Faced with this plethora 
of solutions, some natural questions are whether there is a natural 
measure on the space of solutions, and whether there is
any built-in mechanism in M/string theory that favors some vacua over others.

A possible way of assigning weights to vacua is to consider a Wheeler-deWitt 
equation \cite{DeWitt:1967yk} for families of vacua of string theory. This 
defines a quantum mechanical problem over some space of solutions, and the 
modulus square of the wavefunction is interpreted as a probability 
distribution for the different solutions.

For 4d flux vacua with ${\cal N}=0,1$, the Wheeler-deWitt equation has been 
discussed in \cite{Kobakhidze:2004gm}\footnote{see \cite{Banks:1998vs}
for a discussion of the Wheeler-deWitt equation for M-theory compactifications
preserving 32 and 16 supersymmetries.}. A slightly different scenario for
flux vacua was considered in \cite{Ooguri:2005vr}, where compactifications of
type IIB string theory on $S^1\times S^2\times CY$ were studied. For each 
choice of flux, a wavefunction $\Psi_{(p,q)}$ was introduced, and using the 
intimate connection between flux vacua  and the attractor mechanism, it was 
then argued that the peaks of the different wavefunctions are given by the 
exponentials of the entropies of related black holes with charges $(p,q)$.

In this scenario, it is natural to ask which flux vacuum has a wavefunction
with the largest peak, which semiclassically is equivalent to maximizing
the exponentiated K\"ahler potential
\begin{equation}
S=-\frac{i \pi}{4}\int _M \Omega \wedge \bar \Omega
\end{equation}
over the complex structure moduli space of the Calabi-Yau. However, this 
question is not well posed as stated, since this functional is not invariant
under K\"ahler gauge transformations. Physically, this is related to the fact 
that a rescaling of the charges of the black hole leaves invariant the 
attractor point, but rescales the entropy. To fix this rescaling ambiguity,
the authors of \cite{Gukov:2005bg} propose to extremize this functional 
subject to the condition that we hold fixed one of the periods, 
\begin{equation}
\int _C \Omega =1
\end{equation}
for some 3-cycle $C$. In terms of black holes, this amounts,
via the attractor equations, to scan only over black holes with one of their
charges fixed at a particular value. The entropic principle claims that the 
maximum of this action, subjected to the constraint 
of fixing one period, constitute the preferred Calabi-Yau.

In this way of presenting the principle, the choice of period to be fixed 
seems quite arbitrary. The authors of \cite{Gukov:2005bg}  argue that an 
equivalent formulation of their principle consists in finding the points in 
Calabi-Yau moduli space where all the periods but one are aligned.
To decide if a critical point is a maximum, saddle point or minimum, we need
to compute the second variation of the functional. The character of the 
critical point is then determined by the signature of a reduced 
period matrix, $\hbox{Im } \tau_{ij}$. One might have hoped that in the best 
possible scenario, for a given Calabi-Yau moduli space, of all such critical 
points, only one or very few are actually maxima.

In this note we argue that at the critical points of compact Calabi-Yaus,
the reduced period matrix always satisfies $\hbox{Im } \tau_{ij}> 
0$\footnote{The only possible exception being if $\hbox{Im }\tau_{ij}$ has
one or more zero eigenvalues.}. This implies that all critical points are 
maxima. Since for a given Calabi-Yau moduli space one expects an 
infinite number of critical points (due to the possibility of fixing 
infinitely many different periods), our result shows that the entropic 
principle will need extra input as to why any particular period should 
be fixed, in order to be useful in selecting a single point in Calabi-Yau
moduli space.

In the next section we give a brief review of the construction of wavefunctions
for string vacua, and the entropic principle of \cite{Gukov:2005bg}. We then
proceed to show our main result, namely that all critical points satisfy
$\hbox{Im } \tau_{ij}>0$. A simple family of examples is considered in the
last section.

\section{Wavefunctions for flux vacua and the entropic principle}
In this section we start by recalling the basics of the special geometry
of the complex structure moduli space of a Calabi-Yau \cite{Strominger:1990pd,
Candelas:1990pi}. After reviewing the construction of wavefunctions of
\cite{Ooguri:2005vr} and the proposal of an entropic principle of 
\cite{Gukov:2005bg}, we show that all critical points of this extremization 
problem are actually maxima.

\subsection{Special geometry}
Given a compact Calabi-Yau $M$, we define the Hodge-Riemann bilinear form 
\cite{griffiths} on its cohomology groups, $Q:H^{3-k}\otimes H^{3-k}
\rightarrow \bC$ by
\begin{equation}
Q(\alpha,\beta)=\int _M\alpha \wedge \beta \wedge J^k
\end{equation}
The Hodge-Riemann relations assert that if $\alpha$ is a primitive $(p,q)$ 
form, then\footnote{The overall sign of the Hodge-Riemann bilinear
form fluctuates in the literature, due to different choices of orientation. We 
follow the conventions of \cite{griffiths}.}
\begin{equation}
i^{p-q}(-1)^{\frac{(3-p-q)(2-p-q)}{2}}Q(\alpha,\bar \alpha)>0
\end{equation}
With these conventions, the exponentiated K\"ahler potential 
(multiplied here by a convenient constant) is semipositive
\begin{equation}
S=\frac{\pi}{4}e^{-K_{cs}}=i^3
\frac{\pi}{4}\int_M \Omega \wedge \bar \Omega
\end{equation}
Let $\{A^I, B_I\}$, $I=0,\dots,h^{2,1}(M)$ be a symplectic basis for 
$H_3(M,\bZ)$, and $(\alpha_I, \beta^I)$ their Poincar\'e 
dual forms $(\alpha_I, \beta^I)$. In this basis, the periods are
\begin{equation}
X^I=\int_{A^I}\Omega \hspace{1cm} F_I =\int _{B_I}\Omega
\end{equation}
One can argue that the periods $F_I$ derive from a prepotential $F_I=\partial_I
{\cal F}$, with ${\cal F}(X)$ a homogeneous function of degree two. In terms 
of the periods,
\begin{equation}
S=\frac{\pi}{4}e^{-K_{cs}}=\frac{i\pi}{4}\left\{\bar X^IF_I-X^I\bar F_I\right\}
\end{equation}
Finally, define the period matrix as $\tau_{IJ}=\frac{\partial F_J}
{\partial X^I}$. Since $\partial_I \Omega = \alpha_I-\tau_{IJ}\beta ^J$, it 
follows that 
\begin{equation}
\hbox{Im }\tau_{IJ}=
\frac{i}{2}\int_M\frac{\partial \Omega}{\partial X^I} \wedge 
\overline{\frac{\partial \Omega}
{\partial X^J}}
\end{equation}
Now, using that $\partial_I\Omega \in H^{3,0}\oplus H^{2,1}$, and applying 
the Hodge-Riemann relations we learn that $\hbox{Im }\tau_{IJ}$
has signature $(1,h^{21})$\footnote{The overall sign convention mentioned in 
the previous footnote would show up also here. The result that is 
independent of conventions is that for an exponentiated K\"ahler potential 
that is real and semipositive definite, its matrix of second derivatives 
($-\hbox {Im }\tau_{IJ}$ in our conventions) has signature $(1,h^{21})$, where 
the first entry denotes the number of negative eigenvalues.}. 

The periods provide projective coordinates for the moduli space, but it will
be convenient to introduce affine coordinates $a^i=X^i/X^0$, $i=1,\dots,
h^{2,1}(M)$ and define $F(a)=(X^0)^{-2}{\cal F}(X)$, $a_i^D=
\frac{\partial F(a)}{\partial a^i}$. The action can be 
rewritten in terms of these coordinates as
\begin{equation}
S=\frac{i\pi}{4}|X^0|^2\left \{2(F-\bar F)-(a^i-\bar a^i)(a_i^D+\bar a^D_i)
\right\}
\end{equation}
\subsection{Wavefunctions for compactifications without fluxes.}
There are some obvious similarities between the period matrix $\hbox{Im } 
\tau_{IJ}$ and the deWitt metric: both are defined over a space of space-like 
metrics, and both have hyperbolic signature, with the 'timelike' direction
reflecting the possibility of rescaling the spacelike metrics. It is then
natural to consider whether there is a minisuperspace approach to Calabi-Yau
compactifications, where $\hbox{Im } \tau_{IJ}$ plays the role of metric in
superspace. String theory provides a very natural candidate for a quantum 
mechanical system defined over $H^3(M)$: the B-model topological
string. The first hint is that the operators of the B-model topological string
are in one to one correspondence with the $(2,1)$ cohomology of the Calabi-Yau.
More importantly, it was argued in \cite{Witten:1993ed} that the B-model 
topological string partition function is a wavefunction on $H^3(M)$. This 
partition function satisfies the holomorphic anomaly equations 
\cite{Bershadsky:1993cx}. These holomorphic anomaly equations can be written in
terms of large phase space variables \cite{Dijkgraaf:2002ac,Verlinde:2004ck}, 
with $\hbox{Im } \tau_{IJ}$ playing the role of metric in this large phase 
space. The holomorphic anomaly equations, however, should not be thought of as
the Wheeler-deWitt equation for the topological string: they only reflect the 
background dependence on the choice of polarization\footnote{I would like to
thank R. Dijkgraaf and K. Skenderis for useful comments on this point.}. It 
would be interesting to understand what wave equation plays the role of 
Schr\"odinger equation for the topological string  partition function.

\subsection{Wavefunctions for flux vacua.}
Another family of string compactifications where wavefunctions can be 
computed was discussed in \cite{Ooguri:2005vr}. They consider type IIB with all
spatial dimensions compactified, on $S^1\times S^2\times M$, with $M$ a 
Calabi-Yau, and fluxes turned on. More specifically, a $F_5$ flux of the form 
$w\wedge F_3$ is turned on, where $w$ is a unit form on
$S^2$ and $F_3$ is a RR form determined by the fluxes $(p^I,q_I)$,
\begin{equation}
p^I=\int_{A^I}F_3 \hspace{1cm}q_I=\int_{B^I}F_3
\end{equation} 
The conditions for this flux to preserve supersymmetry are intimately related 
to the attractor equations of a different system: that of the compactification 
on $M$ with wrapped D3-branes with charges $(p^I,q_I)$. The attractor 
mechanism fixes a point in the complex structure moduli of $M$ in terms of the 
charges $(p^I,q_I)$
\begin{equation}
p^I=\hbox{Re }(CX^I)\;\;\;\;\;\; q_I=\hbox{Re }(CF_I)
\end{equation}
For a fixed choice of $(p^I,q_I)$, \cite{Ooguri:2005vr} define the 
entropy functional 
\begin{equation}
S_{(p,q)}=-i\frac{\pi}{4}\left(\int \Omega \wedge \bar \Omega+
\int (\Omega + \bar \Omega)\wedge F_{3}\right)
\label{fluxfunc}
\end{equation}
This functional has two nice properties: first, when we extremize it with
respect to $X^I$, the equations of motion are the attractor equations, 
for $C=1$. Second, when this functional is evaluated at the solution, 
it gives the semiclassical entropy of a black hole of charges $(p^I,q_I)$, 
\begin{equation}
S_{(p,q)}|_{att}=S_{BH}(p^I,q_I)
\end{equation}
The authors of \cite{Ooguri:2005vr} promote the classical BPS equations for the
attractor flows to a supersymmetric version of the Wheeler-deWitt equation. 
In this fashion, for each choice of fluxes $(p,q)$, one has a wavefunction 
$\Psi_{(p,q)}$. These wavefunctions are peaked at the attractor point, and
furthermore it can be argued \cite{Ooguri:2005vr} that their natural 
normalization is such that the value at the peak behaves like 
$e^{S_{BH}(p,q)}$. 

\subsection{The entropic principle}
If we repeat this process for very many 
different charge vectors $(p^I,q_I)$, we will find many attractor points in 
moduli space, and we might be tempted to assign to each of them the weight 
$e^{S_{BH}(p^I,q_I)}$. However, this assignment is not well defined as it 
stands: if we consider a set or 
rescaled charges $(\lambda p^I, \lambda q_I)$, the solution to the attractor 
equations is still the same point in moduli space (recall the $X^I$ are 
projective coordinates), but the entropy of the corresponding black hole 
has changed. If we want to unambiguously  assign a weight to a given point 
in the 
Calabi-Yau moduli space, we first need to fix this ambiguity in the rescaling 
of the charges. A proposal for how to do this was presented in 
\cite{Gukov:2005bg} to which we now turn.

Consider the flux independent part of the entropy functional 
\begin{equation}
S=i^3\frac{\pi}{4}\int _M \Omega \wedge \bar \Omega
\end{equation}
This functional evaluated at an attractor point still gives the semiclassical 
entropy of the corresponding black hole, so the first thought might be trying
to maximize it. However, as we have just seen, as far as determining attractor 
points in moduli space, the charge lattice behaves like a projective variety: 
$(\lambda p^I, \lambda q_I)$ yield the same attractor point that $(p^I,q_I)$. 
A possibility is then to fix a hypersurface in this charge space, {\it e.g.} 
fix one of the  charges. In particular,  in \cite{Gukov:2005bg} they propose 
to extremize the functional $S$, keeping fixed one period $X^c$. Since we are 
fixing the period of a 3-cycle, without loss of generality we take that cycle 
to be primitive. Now, since the action of the symplectic group is transitive, 
we can always take that primitive cycle to be an A-cycle in a symplectic basis.
So without loss of generality, we can fix the period $X^0$. For this 
purpose, one first rewrites this entropy functional in terms of the 
``reduced'' periods $a^i=X^i/X^0$, $a_i^D$. 
\begin{equation}
S=\frac{i\pi}{4}|X^0|^2\left\{2(F-\bar F)-(a^i-\bar a^i)
(a_i^D+\bar a_i^D)\right\}
\end{equation}
At the attractor point, this is 
equivalent to fixing a magnetic charge $p^0$ and $\phi^0$, the chemical 
potential of its dual electric charge\cite{Ooguri:2004zv}. In this way we 
get rid of the ambiguity in assigning an entropy to points in moduli space, 
and comparing the different entropies becomes well-defined.

This entropy functional has critical points with respect to the variation of 
$a^i, \bar a^i$, given by the equation
\begin{equation}
\hbox { Im } a_i^D - \tau_{ij} \hbox { Im } a^j=0
\end{equation}
Since the signature of $\hbox{Im }\tau_{IJ}$ is $(1,h^{2,1})$, the signature 
of $\hbox{Im }\tau_{ij}$ is either $(0,h^{2,1})$ or $(1,h^{2,1}-1)$. 
Furthermore, taking the imaginary part of this equation, we see that if
$|\hbox{Im }\tau_{ij}|\neq 0$, all critical points have $\hbox{Im }a^i=0$.
To determine if these critical points are maxima, minima or saddle points, we 
need to consider the matrix of second derivatives
\begin{equation}
\delta^2S=-\pi|X^0|^2\left(\hbox{Im }\tau_{ij}\right)\delta a^i \delta \bar a^j
+\frac{\pi}{2}|X^0|^2 \hbox{Im }a^i\left(c_{ijk} \delta a^j\delta a^k
+\bar c_{ijk}\delta \bar a^j \delta \bar a^k\right)
\end{equation}
with $c_{ijk}=\partial _i\tau_{jk}$. Since $c_{ijk} \hbox{Im } a^k$ does not 
have a definite signature, if it is different from zero, the critical
point can't be a maximum. A sufficient condition for  $c_{ijk}\hbox{Im }a^k=0$
at a critical point is that $|\hbox{Im }\tau_{ij}|\neq 0$, since then
$\hbox{Im }a^i=0$. Therefore, if $\hbox{Im }\tau_{ij}>0$ the critical point 
is a maximum, and if $\hbox{Im} \tau$ has hyperbolic signature, the critical 
point is a saddle point (or a minimum if $h^{2,1}=1$). Maxima are then of 
the form
\begin{equation}
\hbox{ Im }a^i= \hbox {Im } a^D_i=0\;\;\;\;\; \hbox{Im }\tau_{ij}>0
\end{equation}
Such a point in moduli space solves the attractor equations for $p^i=q_i=0$.
This solution allows for an alternative reformulation of the 
principle \cite{Gukov:2005bg}: the critical points in Calabi-Yau moduli space 
are such that all the periods but one are aligned. Then, the 3-cycle whose
period is fixed is the symplectic dual to the 3-cycle that has null 
intersection with all the 3-cycles with aligned periods.

We now will give two arguments showing that at critical points it is always 
the case that $\hbox{Im }\tau_{ij}\geq 0$, with the equality only possible
if furthermore $\partial _i \Omega=0$ for some $i$. Otherwise, we have a 
maximum. The first argument is quite simple. Using that $X^0$ is 
fixed, we can write the imaginary part of the reduced period matrix as
\begin{equation}
\hbox{Im }\tau_{ij}=\frac{i}{2|X^0|^2} \int_M \frac{\partial \Omega}{\partial 
a^i}\wedge \overline {\frac {\partial \Omega}{\partial a^j}}
\end{equation}
By Griffiths transversality, we know that 
$\partial _i \Omega \in H^{3,0}\oplus H^{2,1}$. 
Specifically \cite{Candelas:1990pi},
\begin{equation}
\frac{\partial \Omega}{\partial a^i}=-\frac{\partial K}{\partial a^i}
\Omega+D_i\Omega
\end{equation}
But since $S=\frac{\pi}{4}e^{-K}$, the equation for critical point 
$\partial _iS=0$ is equivalent to $\partial _iK=0$ unless $K=\pm \infty$. So, 
at critical points $\partial _i \Omega \in H^{2,1}$, and it then follows from 
the Hodge-Riemann relations that at a critical point $\hbox{Im }\tau_{ij}>0$.

We now give a slightly different proof. Starting with the the definition 
$F(a)=(X^0)^{-2}{\cal F}(X)$, we can write the second derivatives of the full
prepotential in terms of reduced quantities
\begin{equation}
\frac{\partial^2 {\cal F}}{\partial X^I \partial X^J}=\left(
\begin{array}{cc}
2F-2a^i\partial_iF+a^i \tau_{ij} a^j & \partial_iF-a^j \tau_{ji} \\
\partial_iF-a^j \tau_{ji} & \tau_{ij}
\end{array}
\right)
\end{equation}
If we now consider the imaginary part of this equation and take determinants,
we arrive at the relation
\begin{equation}
|\hbox{Im }\tau_{IJ}|=-\frac{2S}{\pi |X^0|^2}\left |\hbox{Im }\tau_{ij}+
\frac{2}{\pi |X^0|^2}\frac{\hbox{Re }\partial_iS \hbox {Re }\partial _jS}{S}
\right |
\end{equation}
At a critical point, this yields
\begin{equation}
|\hbox{Im } \tau_{IJ}|_{crit}= -\frac{2S}{\pi |X^0|^2} |\hbox{Im }\tau_{ij}
|_{crit}
\label{chuqui}
\end{equation}
We know that $|\hbox{Im } \tau_{IJ}|<0$ and $S>0$, even away from critical 
points, so it follows that $|\hbox{Im } \tau_{ij}|_{crit}>0 $. Since the only 
two possibilities were that $\hbox{Im }\tau_{ij}>0$ or that 
$\hbox{Im }\tau_{ij}$ had hyperbolic signature, we again conclude that all 
critical points have $\hbox{Im }\tau_{ij}>0$.

Some remarks are in order. First, we have carried out the discussion 
exclusively in the complex structure moduli space. One can adapt the discussion
to the complexified K\"ahler moduli space, and in fact, an equation very 
similar to (\ref{chuqui}) appeared already in \cite{Candelas:1990pi} (see the
last equation in their section 4), valid in the interior of the complexified
K\"ahler cone, and when one fixes the period of the 0-cycle. However, a 
crucial difference is that in that particular case, the reduced period matrix 
has always hyperbolic signature: this is ultimately due to the fact that the
Hodge-Riemann form has definite sign on $H^{2,1}$, while on $H^{1,1}$ has 
hyperbolic signature, since it differentiates the K\"ahler class from 
primitive forms. This does not conflict with mirror symmetry, since mirror 
symmetry between Calabi-Yau manifolds $M$ and $W$ does not exchange 
the Hodge-Riemann form on $H^{2,1}(M)$ with that on $H^{1,1}(W)$.

Secondly, our discussion was restricted to compact Calabi-Yaus. For 
non-compact Calabi-Yaus, all the periods can actually be aligned at some 
locus of the moduli space: this has been studied for orbifolds 
\cite{Douglas:2000qw}, $ALE$ fibrations \cite{Fiol:2000pd}, 
and line bundles over del Pezzo surfaces \cite{Aspinwall:2004vm}. However,  
when zooming in a compact Calabi-Yau into a non-compact one, some periods 
become infinite, and there is no symplectic basis in general. The results we 
derived do not necessarily apply in the non-compact case, but we see this as 
an indication of the limitations of studying non-compact Calabi-Yaus, which at
any rate are not our ultimate interest.

\section{Some examples of critical points}
As an illustration of our results, we will revisit the case of the large 
complex structure limit of one-modulus models, already discussed 
in \cite{Gukov:2005bg}. There, an infinite number of critical points were 
identified, and it was further claimed that there were minima of the entropic
principle. We argue here that, in accordance with our general results, those
critical points are actually maxima. 

The prepotential for these models in the large complex structure limit is
\begin{equation}
{\cal F}=-\frac{1}{3}\frac{(X^1)^3}{X^0}
\end{equation}
In \cite{Gukov:2005bg}, they focused on the subspace with 
$\hbox{Im }X^0=\hbox{Re }X^1=0$, and noted that points where
\begin{equation}
\left(\frac{X^1}{X^0}\right)^2=n
\end{equation}
for some negative integer $n$ are critical points. To ease the notation, it is
convenient to add a quadratic term to the prepotential
\begin{equation}
{\cal F}=-\frac{1}{3}\frac{(X^1)^3}{X^0}+nX^0X^1
\end{equation}
This is equivalent to the $SL(4,\bZ)$ transformation performed 
in \cite{Gukov:2005bg}. The periods are then
\begin{equation}
\left(
\begin{array}{c}
X^0 \\ X^1 \\ F_1 \\ F_0
\end{array} \right)=
\left(
\begin{array}{c}
X^0 \\ X^1 \\
- \frac{(X^1)^2}{X^0}+nX^0 \\ \frac{1}{3}
\frac{(X^1)^3}{(X^0)^2}+nX^1 
\end{array}\right)
\end{equation}
Now, as pointed out in \cite{Gukov:2005bg}, whenever $(X^1/X^0)^2=n$, the $F_1$
period vanishes, so the periods $(X^1,F_1,F_0)$ align, while $X^0$ is not
aligned with them. The non-aligned period is symplectically dual to the fixed 
one \cite{Gukov:2005bg}, so the fixed period is $F_0$, which 
is a magnetic (B-cycle) period with respect to the prepotential. To write down
the action (which of course is symplectically invariant) in terms of reduced 
variables we used that the prepotential is homogeneous in terms of the 
A-periods, and all the subsequent discussion of full vs. reduced period matrix 
is based on a given choice of symplectic basis. Therefore, before we introduce 
reduced periods by fixing $F_0$, we perform a symplectic transformation so 
$F_0$ becomes an electric (A-cycle) period,
\begin{equation}
\left(
\begin{array}{c}
\tilde {X^0} \\ \tilde {X^1} \\ \tilde {F_1} \\ \tilde {F_0}
\end{array} \right)=
\left(
\begin{array}{c}
F_0 \\ F_1 \\
-X^1 \\ -X^0 
\end{array}\right)
\end{equation}
and the new prepotential is $\tilde {\cal F}={\cal F}-(X^IF_I)=-{\cal F}$, 
which differs by a minus sign from the one used 
in \cite{Gukov:2005bg}. In these dual variables we can now define the usual 
reduced variables, starting with $\tilde F=(\tilde X^0)^{-2}\tilde {\cal F}$. 
Straightforward  computation at a critical point $(X^1/X^0)^2=n$ then yields
\begin{equation}
\hbox{Im }\tilde \tau = \frac{1}{4\sqrt{|n|}} >0
\end{equation}
so these points are maxima, as they had to be according to our general 
arguments.

{\bf Acknowledgments}. I would like to thank Sergei Gukov, Kirill Saraikin 
and Cumrun Vafa for correspondence on \cite{Gukov:2005bg} and the 
present paper. I have benefited from insightful comments by Jan de Boer, 
Robbert Dijkgraaf, Lotte Hollands and Kostas Skenderis. Special 
thanks to Erik Verlinde for many discussions and explanations on these 
matters. I would also like to thank the theory groups at the Universidad 
Aut\'onoma de Madrid and the Universitat de Barcelona for giving me the chance 
to present related material at their respective Xmas. Workshops. Finally, 
during the course of this work, I greatly enjoyed the hospitality of the 
theory groups (and the pastry chef) at CERN and at the Universidade de 
Santiago de Compostela.

\end{document}